\documentclass[paper,twocolumn,showpacs,superscriptaddress,prl]{revtex4}%
\usepackage{amsfonts}
\usepackage{amsmath}
\usepackage{amssymb}
\usepackage{graphicx}%
\begin{document}
\title{The momentum-resolved and time-resolved two-color optical coherence absorption spectrum in the scattering process}
\author{Jiang-Tao Liu}
\email[Electronic address:]{jtliu@semi.ac.cn}
\affiliation{Department of Physics, Nanchang University, Nanchang
330031, China}%
\author{Fu-Hai Su}
\affiliation{Key Laboratory of Materials Physics, Institute of Solid
State Physics, Chinese Academy of Sciences, Hefei 230031, China}
\author{Xin-Hua Deng}
\affiliation{Department of Physics, Nanchang University, Nanchang
330031, China}%
\author{Hai Wang}
\affiliation{Department of Physics, Capital Normal University,
Beijing 100037, China}%
\pacs{42.65.-k, 72.25.Rb, 78.40.-q}

\date{\today}

\begin{abstract}
The two-color optical coherence absorption spectrum (QUIC-AB) of GaAs quantum well in the presence of a charge current is investigated. We find that the QUIC-AB depends strongly not only on the amplitude of the electron current but also on the direction of the electron current. Thus, the amplitude and the angular distribution of scattering current in the scatter process can be detected directly in real time with the QUIC-AB. The phase shift of scattered waves and the details of the scattering potential can also be determined.
\end{abstract}
\maketitle

%\altaffiliation{Author to whom correspondence should be addressed. Electronic address: kchang@red.semi.ac.cn}
%\affiliation{NLSM, Institute of Semiconductors, Chinese Academy of Sciences, P. O. Box 912,
%Beijing 100083, China}

Scattering mechanisms play a very important role in the electron and spin transport. For instance, the extrinsic spin Hall effect is mainly attributed to the skew-scattering and side-jump mechanism \cite{1MI,2LB,3JE,4SZ}. However, in most of these studies, the contribution of the scattering interaction was investigated based on the non-equilibrium statistical mechanics (NESM) \cite{2LB,4SZ,5EM}. The NESM always involve the multiple scattering process, which makes the systems too complex to obtain the detail interactions of carries and the scattering potential. Thus, is there a way to measure the interactions of carries and scattering potential directly? The collision experiment demonstrates high efficiency in the study of particle-particle interaction. In the particle collision experiment, when the angular distribution of scattering particles is detected, the phase shift of the scattered waves $\delta_{l}$ and the details of the particle-particle interaction is obtained.

However, to realize the collision experiment in semiconductors, some problems must be solved. First, high-speed carriers should be injected efficiently, such as, by employing the two-color quantum interference control (QUIC) techniques \cite{6RD, 7MJ, 8JH}. In the QUIC, the carriers can generate at a speed of ~1000 km/s \cite{9HZ, 10LK}. Second, the collision between the injected carriers and the scattering potential should not be more than one time, which is easy to realize in the particle collision experiment by using a very thin target. However, for traditional transport processes in semiconductors, e.g., the spin Hall effect, the collisions between carriers and impurities are frequent because of the long propagation distance of the injected carriers. Fortunately, benefiting from the development of the pump-probe technique, the time resolution can be less than 100 fs. Thus, in an appropriate time interval ($\sim200$ fs), the propagation distance of the injected carriers becomes so short ($\sim200$ nm) that the collisions between carriers and impurities can be less than one time. Finally, to obtain the phase shift of the scattered waves, both the amplitude and the angular distribution of scattering current must be detected. Although some experimental techniques have been used to observe the distribution of the electrons in the momentum space (e.g., angle-resolved photoemission spectroscopy) \cite{11SH, 12AD}, only a few of these techniques have high time resolution.

\begin{figure}[b]
\includegraphics[width=0.9\columnwidth,clip]{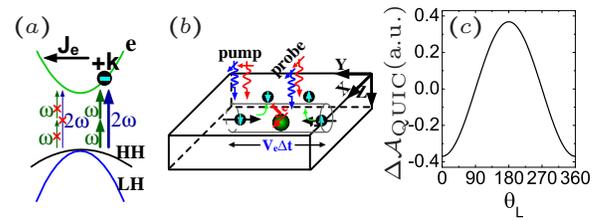}
\caption{(color online). (a) Schematic illustration of the transition of two parallel-linearly polarized beams in semiconductor materials with a flowing current.
(b) Illustration of cross-linearly polarized $\omega$ and $2\omega$ pulses producing a pure spin current along the $\widehat{y}$ direction. A pure charge current is injected through the spin-dependent asymmetric scattering processes. The angular distribution of scattering pure charge current is detected by  parallel polarized $\omega$ and $2\omega$ light beams. (c) The differential absorption of QUIC-AB as a function of the light polarization direction, with electric current $J_{e}=10^{5}$ A/cm$^{2}$ flows along the $-\mathbf{k}$ direction.}%
\label{fig1}%
\end{figure}

In this Letter, we suggest a different scheme by utilizing a momentum-resolved two-color optical coherence absorption spectrum (QUIC-AB) for the direct detection of scattering current in real time. In QUIC, the optical transition rate depends strongly on the direction of the wave vector $\mathbf{k}$ \cite{6RD, 7MJ, 8JH, 13AN}; e.g.,  the transition rate of QUIC can be enhanced at $+\mathbf{k}$ but disappears at $-\mathbf{k}$. Thus, if the band states near $+\mathbf{k}$ are already filled with electrons [see Fig. 1(a)], the optical transition rate of QUIC-AB is reduced due to the state filling effects.  However, if the electrons are located at $-\mathbf{k}$, these electrons have no effect on the QUIC-AB because the optical transition at $-\mathbf{k}$ is forbidden.

To reveal the relation between the electron distribution function and the QUIC-AB, we calculate the transition rate of QUIC using the Volkow-type wave function \cite{14AS, 15JT, 16JT}. In optical QUIC, the total vector potential of $\omega$ and $2\omega$ laser pulse is given by $\mathbf{A}=\mathbf{a}_{1}A_{1}\cos(\omega t+\varphi_{1}) +\mathbf{a}_{2}A_{2}\cos(2\omega t+\varphi_{2})$. The transition rate of QUIC can be written as
\begin{align}
W(\mathbf{k})=2\pi\left(\frac{e_{0}}{\hbar2m_{0}c_{0}}\right)^{2} \delta\left[\omega_{cv}\left(  \mathbf{k}\right)  -2\omega\right]\ \ \ \ \ \  \ \nonumber\\
\times \left(  W_{\mathbf{k}}%
^{n}+W_{\mathbf{k}}^{i}\right)[f_{v}(\mathbf{k})-f_{c}(\mathbf{k})],\ \ \label{trans_rate}
\end{align}
where
\begin{align}&W_{\mathbf{k}}^{n}=\left(\mathbf{k}\cdot\mathbf{a}
_{1}\right)  ^{2}\left\vert
\mathbf{p}_{vc}\cdot\mathbf{a}_{1}\right\vert
^{2}\eta_{1}^{2}A_{1}^{2}+\left\vert
\mathbf{p}_{vc}\cdot\mathbf{a}_{2}\right\vert ^{2}A_{2}^{2},\nonumber\\
&W_{\mathbf{k}}^{i}=\mathbf{k}\cdot\mathbf{a}_{1}\eta_{1}A_{1}A_{2}%
\left[ \left(  \mathbf{p}_{vc}\cdot\mathbf{a}_{1}\right)
^{\ast}\left(\mathbf{p}_{vc}\cdot\mathbf{a}_{2}\right)  e^{i\Delta\varphi }+c.c.\right],\nonumber \end{align}
where $m_{0}$ is the electron mass, $e_{0}$ the electron charge magnitude, $c_{0}$ the speed of light in vacuum, $\omega _{cv}$ the optical transition
frequency, $f_{v}(\mathbf{k})$ ($f_{c}(\mathbf{k})$) the distribution function of the valence (conduction) electrons, $\eta_{1}=\frac{e_{0}A_{1}}{2\omega c_{0}
m_{cv}}$, $1/m_{cv}=1/m_{c}-1/m_{v}$, $m_{c}$ ($m_{v}$) the
effective mass of electrons (holes), $\mathbf{p}_{vc}=\langle
c|\mathbf{p}|v\rangle$ the dipole transition matrix element, and $\Delta\varphi=2\varphi_{1}-\varphi_{2}$ the relative phase of the $\omega$ and $2\omega$ laser beams. $W_{\mathbf{k}}^{n}$ and $W_{\mathbf{k}}^{i}$ describe the single-photon and two-photon transition and the quantum interference between the two-color photon excitations. Thus, the optical absorption coefficient is given by $\alpha(\omega)=\frac{\hbar \omega W(\omega)}{u(c_{0}/n_{0})}$, where $u=\frac{n_{0}^2A_{2}^{2}\omega^{2}}{2 \pi c_{0}^{2}}$ is the electromagnetic energy density, and $n_{0}$ the refractive index of the semiconductors.

From Eq. (\ref{trans_rate}), only the interference term $W_{\mathbf{k}}^{i}$ depends on both the direction and the value of electron vector $\mathbf{k}$.  In the current paper, we focus on the detection of pure electron current. Thus, two parallel linearly polarized probe light (i.e., $\mathbf{a}_{1}\|\mathbf{a}_{2}$) are used. In this case, the interference term $W_{\mathbf{k}}^{i}$ reaches the maximum for $\Delta\varphi=0^{\circ}$ but disappears for $\Delta\varphi=90^{\circ}$. The normal single-photon and two-photon transition term $W_{\mathbf{k}}^{n}$ is invariant for different $\Delta\varphi$. If we measure the difference in absorption between $\Delta\varphi=0^{\circ}$ and $\Delta\varphi=90^{\circ}$, the normal single-photon and two-photon transition term $W_{\mathbf{k}}^{n}$ can be eliminated.  Thus we can define the absorption coefficient of QUIC-AB as%
\begin{equation}\alpha_{\text{\tiny QUIC}}\left(  \omega\right)  =\alpha_{\Delta\varphi=0}\left(\omega\right) -\alpha_{\Delta\varphi=90^{\circ}} (\omega).\end{equation} %
Using this definition, the noises induced by other absorption process (e.g., the intraband absorption, the heating effect, and the electro-optic effect) can also be eliminated and  the precision of measurement can be improved.

 Using Eq. (\ref{trans_rate}), the absorption coefficient of QUIC-AB in the presence of a  charge current can be written as
\begin{align}
\alpha_{\text{\tiny QUIC}}\left(  \omega\right)  =\frac{\pi\omega n_{0}e_{0}^{2}}{2\hbar
uc_{0}^{3}m_{0}^{2}}\eta_{1}A_{1}^{2}A_{2}P^{2} \ \ \ \  \ \ \ \  \ \ \ \  \  \ \  \ \nonumber\\
\times\underset{\mathbf{k}}{\sum}\left[  f_{v}(\mathbf{k})-f_{c}%
(\mathbf{k})\right]  \delta\left[  \omega_{cv}\left(  k\right)  -2\omega
\right]k\cos\theta,\label{QUIC-AB}
\end{align}
where $P^{2}=|\langle S|P_{x}|X|\rangle|^2$, $|S\rangle$ and $|X\rangle$
are  the Kohn-Luttinger amplitudes, and $\theta$ denotes the angle between the electron wave vector $\mathbf{k}$ and the polarization direction of light $\mathbf{a}_{l}$. Thus, the difference in absorption  between that with and that without an electron current can be written as
$\Re\left(  \omega,\Delta\varphi\right)  =\frac{e^{-\alpha\left(
\omega,\Delta\varphi\right)  L}-e^{-\alpha^{0}\left(  \omega,\Delta
\varphi\right)  L}}{e^{-\alpha^{0}\left(  \omega,\Delta\varphi\right)  }%
}\approx\left[  \alpha^{0}\left(  \omega,\Delta\varphi\right)  -\alpha\left(
\omega,\Delta\varphi\right)  \right]  L$
, where $L$ is the thickness of the sample, and $\alpha \left(  \omega,\Delta\varphi\right)$ [$\alpha^{0}\left(  \omega,\Delta\varphi\right)$] is the corresponding absorption coefficient with [without] a electron current. Thus, the differential absorptivity of QUIC-AB  is
\begin{align}
\Delta \mathcal{A}_{\text{\tiny QUIC}}(\omega)&=\Re_{\Delta\varphi=0}(\omega)-\Re_{\Delta\varphi=90^{\circ}}(\omega)\nonumber\\ &\approx[\alpha_{\text{\tiny QUIC}}^{0}(\omega)-\alpha_{\text{\tiny QUIC}}(\omega)]L. \label{QUIC_abL}
\end{align}

As the momentum relaxation time of holes is usually less than 100 fs \cite{a1DJ},  the electrons in the  valence-band follow a uniform angular distribution in k-space (i.e., $f_{v}(\mathbf{k})\equiv f_{v}(k)$). Thus, from Eq. (\ref{QUIC-AB}),  $\alpha_{\text{\tiny QUIC}}^{0}-\alpha_{\text{\tiny QUIC}} \sim \frac{1}{2}\sum_{\mathbf{k}}\left[  f_{c}(-\mathbf{k})-f_{c}(\mathbf{k})\right]\delta_{\omega_{cv},2\omega}  k\cos\theta $,   which is linearly proportional to the current density along the light polarization direction $J_{e}^{\mathbf{a}_{l}}=-e_{0}{\sum}n_{e}(\mathbf{k})\mathbf{v}_{e}(\mathbf{k})\bullet\mathbf{a}_{l}\sim -\frac{\hbar e_{0}}{m_{e}}\sum_{\mathbf{k}}\left[  f_{c}(\mathbf{k}%
)-f_{c}(-\mathbf{k})\right] k\cos\theta$.  Therefore,
 using the  QUIC-AB, the direction and amplitude of the electron  current can be detected directly.

A numerical simulation is shown in Fig. 1(c). In this example, some electrons are located at $+\mathbf{k}$, and the electric current $J_{e}=10^{5}$ A/cm$^{2}$ flows along the $-\mathbf{k}$ direction [see Fig. 1(a)]. When the polarization direction of lights is anti-parallel to the electric current  direction, i.e., $\theta=0^{\circ}$, the absorption in QUIC-AB decreases. When the polarization direction of light is  parallel to the electric current  direction, i.e., $\theta=180^{\circ}$,  these
electrons have no effect on the QUIC-AB although there are some electrons at $+\mathbf{k}$, because the optical transition at $+\mathbf{k}$ is quite small for the quantum interference effect. The optical transition rate at $-\mathbf{k}$ is enhanced at about one time, the absorption in QUIC-AB increases.  Thus, by varying the polarization direction of the probe lights, the angular distribution of the scattering current can be determined.

Benefitting from the pump-probe technique, a high time resolution can be achieved in the QUIC-AB. Thus, the QUIC-AB can provide a different practical scheme for the direct investigation of the scatter processes. In this Letter, we investigate a spin-dependent asymmetric scattering processes similar to the experiment reported by Zhao et. al. \cite{9HZ}. As shown in Fig. 1(b), a pure spin current is generated by the orthogonally polarized $\omega$ and $2\omega$ pulses. After a short time, through asymmetric scattering, such as the skew-jump and the side-jump, a charge current is injected as the spin-down and spin-up electrons are sent in the same direction along $-\widehat{x}$ \cite{1MI, 2LB, 9HZ}. The amplitude and the angular distribution of the scattering current can be detected using the QUIC-AB.  The sample, consisting of 10 periods of $14$ $nm$ wide $Al_{0.1}Ga_{0.9}As$ wells and $14$ $nm$ thick $Al_{0.4}Ga_{0.6}As$ barriers, and the parameters of the pump beams are chosen as follows: full width at half maximum (FWHM) of the pump pulse  $\Delta \Gamma=70$ $fs$, center wavelength of $\omega$ ($2\omega$) lights $\lambda_{\omega}=1430$ $nm$ ($\lambda_{2\omega}=715$ $nm$), and energy density of the $2\omega$ laser pulse $I_{2\omega}=1.5$ $\mu J/cm^{2}$. We tune the $\omega$ pulse fluence to obtain equal transition rates between $\omega$ and $2\omega$ laser pulses.

\begin{figure}[t]
\includegraphics[width=0.98\columnwidth,clip]{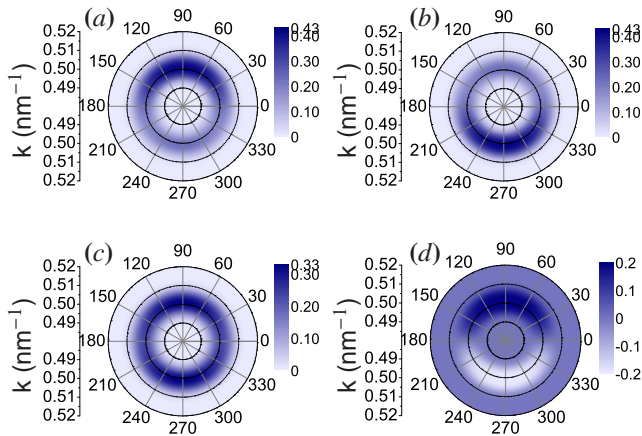}
\caption{(color online). (a), (b) and (c) show the polar contour plot of the pumped spin-up electron, spin-down electron, and total electron distribution function in k-space, respectively. (d) the polar contour plot of the spin current distribution $f_{\uparrow}(+\mathbf{k})$-$f_{\uparrow}(-\mathbf{k})$ in k-space .}
\label{fig2}%
\end{figure}

The electron and spin distribution after pump excitation can also be calculated using Eq. (\ref{trans_rate}). Figs. 2(a), 2(b), and 2(c) describe the spin-up, spin-down, and electron distribution function in k-space, respectively. After pump excitation, the spin-up electrons are concentrated at $\theta_{e}=90^{\circ}$ (i.e., positive $k_{y}$  direction) and vice versa for the spin-down electrons. A pure spin current is injected along the $\widehat{y}$ direction [see Fig. \ref{fig2}(d)].  However, the total electron distribution is still centrally symmetric in the k-space [see Fig. \ref{fig2}(c)]. There is no electron current immediately after pump excitation; i.e., at t=0, the QUIC-AB disappears.

When the pure spin current is injected, a transverse Hall pure charge current is generated by the asymmetric scattering. The scattering of the pure spin current may be caused by different contributions, including the carrier-carrier scattering \cite{17MT}, carrier-phonon scattering \cite{18JM}, skew scattering\cite{5EM, 19JS}, side-jump mechanism \cite{2LB},  and Rashba and Dresselhaus effect \cite{20JS,21SM}. In these mechanisms, the average carrier-phonon scattering relaxation time is usually larger than 1 ps; for the electron-electron scattering, no charge current is injected because the total momentum is conserved; the side-jump mechanism also has little effect on the QUIC-AB because it does not change the scattering angular distribution; Different with Ref. \cite{10LK}, $Al_{0.1}Ga_{0.9}As$ quantum wells is used in our calculation. Thus, the Rashba and Dresselhaus effect, i.e., the intrinsic spin Hall effect, can also be neglected because it is usually quite small in relatively wide band gap and doped semiconductors \cite{5EM, 22WY}. Therefore, the main issue to consider is the skew scattering.

Skew scattering depends on the spin polarization of injected carriers. For example, while the spin up (spin down) electrons  are injected in the flow along the the $+\mathbf{y}$ ($-\mathbf{y}$ ) direction, these electrons are more likely to be scattered in the $-\mathbf{x}$ direction. They will generate a pure charge current in the $+\mathbf{x}$ direction [see Fig .1(b)]. In skew scattering, the scattering potential is given by $V=V_{c} (r)+V_{s}(r)\mathbf{\sigma}\bullet \mathbf{L}$,
where $V_{c}$ and $V_{r}$ are the central potential and spin orbit potential, respectively, and $\mathbf{\sigma}$ ($\mathbf{L}$) is the spin (orbital) angular momentum  of the electron. In the current paper, a circular well potential is used, i.e., $V_{c} (r)=V_{0}H (a_{0}-r)$, $V_{s}(r)=\alpha \hbar \delta(r-a_{0})V_{0}/a_{0}$, where $\alpha$ is the effective spin-orbit coupling constant, and $a_{0}$ the imprity radius,   for $r<a_{0}$, $H (a_{0}-r)=1$, and for $r>a_{0}$, $H (a_{0}-r)=0$. Using the general scattering theory \cite{a2ML}, the scattering rate from $\mathbf{k}$ to $\mathbf{k}'$ is
\begin{equation}
\mathcal{T}^{\sigma}_{k,\vartheta}=|A_{k,\vartheta}|^{2}+|B_{k,\vartheta}|^{2}\sin^{2} \vartheta+2\sigma \sin\vartheta\text{Re}(A_{k,\vartheta}B_{k,\vartheta}^{*}) ,
\end{equation}
where $A_{k,\vartheta}$ and $B_{k,\vartheta}$ are the complex scattering amplitudes, $k$ the magnitude of the vectors of $\mathbf{k}$ and $\mathbf{k}'$, and $\vartheta$ the angle between $\mathbf{k}$ and $\mathbf{k}'$.

Before calculating the scattering current, the collision time should be discussed. To realize the collision experiment in semiconductors, choosing an appropriate time delay $\Delta t$ between the pump and probe lights is important. If the time delay $\Delta t$ is too long, the collision between the injected carriers and the scattering potential becomes more than one time, making the system too complex to analyze at the micro-level. However, in the collision experiment, we also hope that the injected carriers and scattered carriers are far from the impurity; i.e., the propagation distance of the injected carriers should be much larger than the effective range of the impurity potential. Thus, the time delay $\Delta t$ should not be too short. As the velocity of the injected carriers is about 10$^{6}$ m/s, $\Delta t$ is appropriate in the range of about 100-500 fs. In the current paper, we choose $\Delta t=200$ fs.

Before probe lights arrived, the injected electrons are scattered only when an impurity is present in the propagation path, i.e., in the area that equals to $2a_{\text{eff}}v_{e}\Delta t$, where $a_{\text{eff}}$ is the effective range of the impurity potential, and $v_{e}$ is the speed of the injected carriers [see Fig. 1(b)]. Thus, the collision probability between the electron and the impurity potential $\gamma=N_{i}2a_{\text{eff}}v_{e}\Delta t$. At $t=0$, the electron distribution is $\xi_{\sigma,k}(\theta_{e})$, and the electron distribution $\xi'_{\sigma,k}(\theta'_{e})$ at $t=\Delta t$ is given by
\begin{equation}
\xi'_{\sigma,k}(\theta'_{e})=\xi'_{\sigma,k}(\theta'_{e})+\underset{\theta_{e}}\sum \gamma \mathcal{T}^{\sigma}_{k,\vartheta}[\xi_{\sigma,k}(\theta_{e})-\xi_{\sigma,k}(\theta'_{e})]d\theta_{e}.
\label{jk}\end{equation}
Thus, we can define the current density distribution function in k space as $\mathcal{J}_{k,\theta'_{e}}=-\frac{e_{0}\hbar k}{m_{c}}[\xi'_{k}(\theta'_{e})-\xi'_{k}(\theta'_{e}+\pi)]$. $\mathcal{J}_{k,\theta'_{e}}$ corresponds to the asymmetric charge distribution in k-space. As the total electron distribution is symmetrical at t=0,  the asymmetric charge distribution in k-space at $t=\Delta t$ is caused by the spin-dependent asymmetric scattering process, i.e., the skew scattering. As shown in Eqs. (\ref{QUIC-AB}) and  (\ref{QUIC_abL}), we can detect $\mathcal{J}_{k,\theta'_{e}}$ using the QUIC-AB and then obtain the details of the skew scattering.

\begin{figure}[t]
\includegraphics[width=0.75\columnwidth,clip]{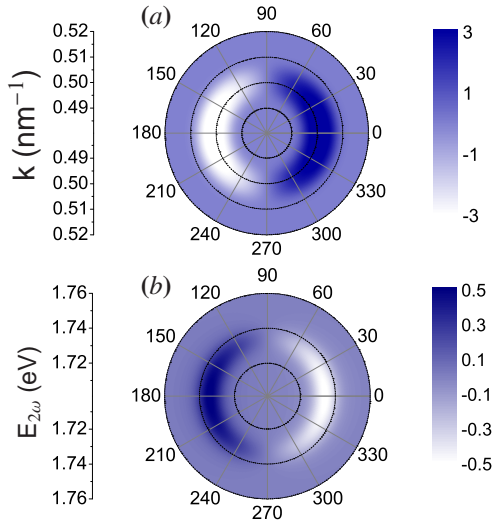}
\caption{(color online). (a) Polar contour plot of the current density distribution function in k space $\mathcal{J}_{k,\theta}$ for $\Delta t=200$ fs, $\alpha\hbar=4.4{\AA}^{2}$, $V_{0}=-5$ meV, and $a_{0}=9.45$ nm. (b) Polar contour plot of QUIC-AB $\Delta \mathcal{A}_{\text{\tiny QUIC}}\times 10^{4}$ as a function of the energy $E_{2\omega}$ and the polarization direction of probe lights $\mathbf{a}_{l}$. }
\label{fig3}%
\end{figure}

The numerical results are shown in Fig. 3. As the injected pure spin current is  along the $+\widehat{y}$ direction, the scattered pure charge current is along the $-\widehat{x}$ direction. Thus the total electron distribution $\xi'_{k,\uparrow}+\xi'_{k,\downarrow}$ at $\theta_{e}=0^{\circ}$ is smaller than that at $\theta_{e}=180^{\circ}$. The current density distribution function in k space $\mathcal{J}_{k,\theta'_{e}}$ around $+k_{x}$ ($-k_{x}$) is positive (negative) [see Fig. 3(a)]. If the polarization direction of the probe lights is also along the $+\widehat{x}$ direction, the absorption of the probe lights is reduced due to the state filling effect;  a negative $\Delta \mathcal{A}_{\text{\tiny QUIC}}$ is then achieved. When the polarization direction of the probe lights deviates from the $+\widehat{x}$ direction, $\Delta \mathcal{A}_{\text{\tiny QUIC}}$ increases.  The QUIC-AB disappears when the polarization direction of probe lights is perpendicular to the scattered pure spin current, i.e., $\theta_{\mathbf{a}_{l}}=90^{\circ}$. When the polarization direction of the probe lights is antiparallel with the charge current, a maximum $\Delta \mathcal{A}_{\text{\tiny QUIC}}\approx 0.50 \times 10^{-4}$ is achieved. Thus, we can detect the angular distribution of the scattering current by varying the the polarization direction $\mathbf{a}_{l}$. If the angular distribution of the scattering current is detected, the phase shift of the scattered waves and the details of the scattering potential can be determined \cite{a2ML}.

The QUIC-AB may have future applications in the study of the ultrafast transport process and ultrafast electronics; e.g.,  a 1THz graphene transistor may be achieved \cite{23KS,24YM}, and thus a different current detection with sub a ps time resolution is very important.

Finally, we discuss the experimental realization of our theory predication.  Benefitted from ultrafast laser technology,  the time resolution of the pump-probe detection can be down to the femto-second. However, the pump and probe pulse length should not be too short. A short pulse with a too large energy range will lower the momentum resolution and reduce the interference term $W_{\mathbf{k}}^{i}$. Thus, the FWHM of the laser pulse is appropriate in the range of about 50-200 fs. As the probe pulse also injects a pure charge current, a relatively low energy density $2\omega$ probe pulse must be chosen. The skew scattering is weak in a short time; Thus, the maximum QUIC-AB is about $0.50 \times 10^{-4}$ [see Fig. \ref{fig3}(b)].  However, it is not difficult to detect using the current technique; e.g., the resolution of the differential transmission $\Delta T/T$ in the experiment can be smaller than $10^{-6}$ \cite{9HZ,10LK}. In the detection of QUIC-AB, there is no spin or charge accumulation in real space. Thus, laser beams need not be focused, and  the disturbance of edge states  can be avoided.

In conclusion, the momentum-resolved and time-resolved QUIC-AB is investigated. The QUIC-AB can be used to detect both the amplitude and the direction of the scattering charge current in the scatter process. The phase shift of scattered waves and the details of the scattering potential can be obtained. The QUIC-AB may carry out the collision experiment in semiconductors  and can play an important
role in investigating the ultrafast transport process or ultrafast electronics and spintronics.

We would like to thank Hui Zhao for fruitful discussion. This work was supported by the NSFC Grant Nos. 10904059 and 10904097.

\end{document}